\documentclass[pre,aps,12pt]{revtex4-1}
\usepackage{graphicx}
\graphicspath{{FIGURES//}}
\usepackage{multirow}
\usepackage{amsmath}
\usepackage{amsfonts}
\usepackage[latin1]{inputenc}

\begin{document}

\title[Routing dynamics]{Criticality in conserved 
       dynamical systems: Experimental observation 
       vs.\ exact properties
       }

\author{Dimitrije Markovi\'c}
\affiliation{Institute for Theoretical Physics,
             Johann Wolfgang Goethe University,
             Frankfurt am Main, Germany}

\author{Andr\'e Schuelein}
\affiliation{Institute for Theoretical Physics,
             Goethe University Frankfurt, Germany}

\author{Claudius Gros}
\affiliation{Institute for Theoretical Physics,
             Johann Wolfgang Goethe University,
             Frankfurt am Main, Germany}

\date{\today}

\begin{abstract}
Conserved dynamical systems are generally considered 
to be critical. We study a class of critical routing models,
equivalent to random maps, which can be solved rigorously
in the thermodynamic limit. The information flow is conserved 
for these routing models and governed by cyclic attractors.
We consider two classes of information flow,
Markovian routing without memory and vertex routing 
involving a one-step routing memory. Investigating 
the respective cycle length distributions for complete graphs
we find log corrections to power-law 
scaling for the mean cycle length, as a function 
of the number of vertices, and a sub-polynomial growth
for the overall number of cycles. 

When observing experimentally a real-world dynamical
system one normally samples stochastically its phase space.
The number and the length of the attractors are then
weighted by the size of their respective 
basins of attraction. This situation is equivalent
to `on the fly' generation of routing tables for which
we find power law scaling for the weighted average 
length of attractors, for both conserved routing 
models. These results show that critical dynamical systems are 
generically not scale-invariant, but may show power-law 
scaling when sampled stochastically. It is hence important
to distinguish between intrinsic properties of a critical
dynamical system and its behavior that one would observe when
randomly probing its phase space.

\end{abstract}

\pacs{05.10.-a,89.75.Da, 89.75.Fb}

\maketitle

{\bf
Power law scaling is observed in many real-world
systems, like the distribution of neural avalanches
in the brain. In statistical physics all critical
systems, at the point of a second-order phase transition,
show power law scaling. Power law scaling is hence commonly 
attributed to criticality, but it is an open question
to which extend this relation is satisfied
for complex dynamical systems. There is, in addition,
a difference between the distribution an observer may
be able to sample and the exact properties of the
underlying dynamical system. An observer will sample in general 
the number and the size of attractors as weighted by size 
of their respective basins of attraction. Here we investigate
critical models for information routing and show that
the number and the length of attractors does not obey 
power law scaling, while, on the other hand, an external 
observer, sampling the weighted distribution, would find 
power law scaling. We hence conclude that drawing
conclusions from experimentally observed power law scaling
needs to take into account the implicitly employed sampling 
procedures.
}


\centerline{\rule{0.6\textwidth}{0.5pt}}

\section{Introduction}

The propagation of perturbations is a central
notion in dynamical system theory.
One speaks of a frozen state when a 
perturbation tends to die out, on the average, during
the course of time evolution and of a 
chaotic state when perturbations tend to spread
out~\cite{derrida86,gros08}. A given class of
dynamical systems may change from frozen to chaotic
behavior as a function of parameters, being
critical right at the transition point.

At criticality, information is on the average conserved \cite{tsuchiya00},
as one can regard a perturbation of a state as the
information about the persistence of small differences. 
A well studied example of a critical dynamical system is 
the Kauffman net with connectivity $K=2$, an example of 
a random Boolean network~\cite{kauffman69,aldana03,drossel08}. 
In statistical mechanics critical systems are generically 
scale invariant~\cite{reichl09}, and it has been widely 
assumed that this statement would also hold for critical 
dynamical systems. Indeed numerical simulations seemed to 
support scaling in critical Boolean networks, notably a 
$\sqrt{N}$ scaling for the number of attractors as a 
function of the number of vertices $N$ had been 
proposed~\cite{kauffman69,aldana03}.

An important clarification then came with the exact proof
that the number of attractors actually grows faster than
any power of $N$, and that the results of the numerical simulations
suffered from systematic undersampling of phase space~\cite{samuelsson03}.
It could be shown, on the other side, that the number of
frozen and the number of relevant nodes in a large class 
of critical Boolean networks obey power law scaling~\cite{mihaljev06}.
The situation is then that certain properties of critical
dynamical systems, at least for the case of random Boolean
networks, obey power law scaling while others do not. 
It is hence important to investigate the possible 
occurrence of scaling in different classes of dynamical 
systems. 

We study a class of dynamical systems describing 
the transport of conserved quantities on network structures, 
that is quantities which cannot be multiplied or 
separated into smaller parts during the transport between 
network nodes. We denote such a process a routing process, 
since only one node is active at each time step, the one 
containing the transmitted quantity.
A routing process can be seen alternatively as the 
transport of  perturbations between network elements 
and as such represents a critical process because the perturbation 
neither spreads out through the entire network nor does it die out. 
A routing process initiated from a given network node will eventually 
follow a limiting cycle, thus the total number of nodes 
affected by the perturbation will be a finite fraction 
of the whole network. Hence, a routing process satisfies 
the conditions needed for it to be considered as a critical 
dynamical process~\cite{luque2000}.

\begin{figure}[tb]
\centerline{
\includegraphics[width=0.9\textwidth]{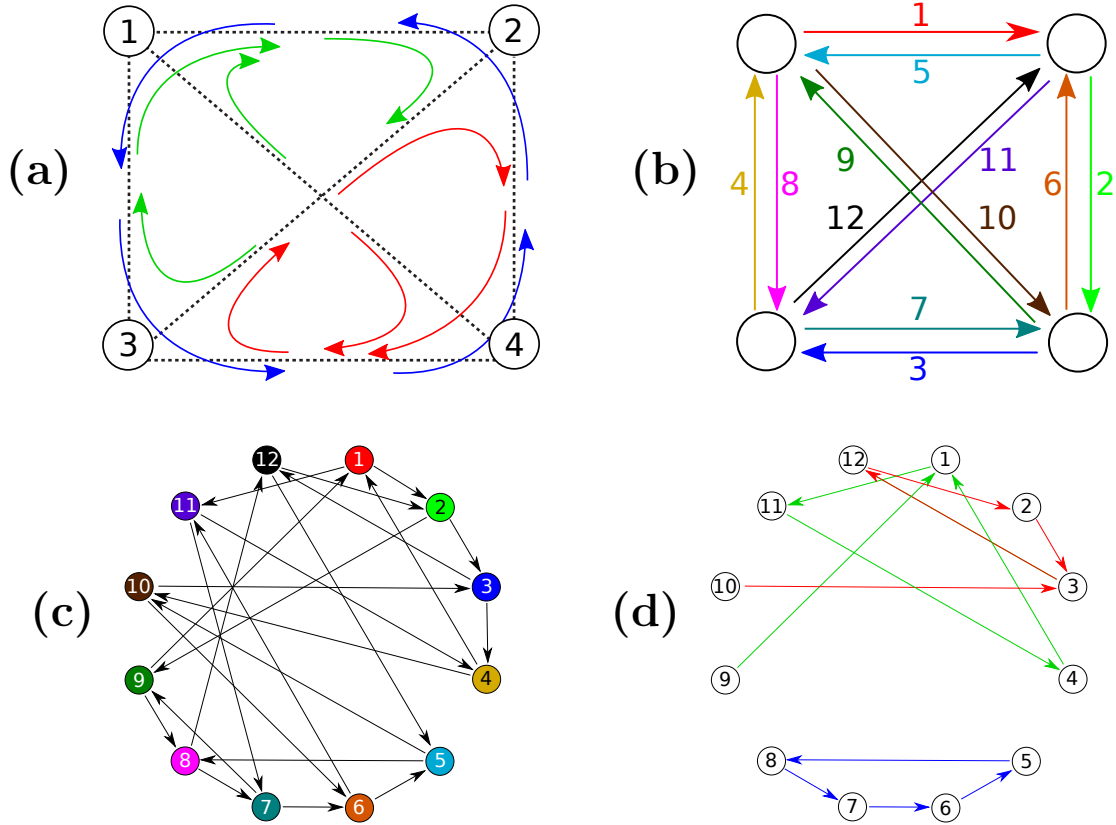}
           }
\caption{Vertex routing dynamics for a $N=4$ complete graph
{\bf (a)} A realization of the routing tables.
Routing through the first vertex follows
$T_{312}=T_{213}=T_{214}=1$, with all other $T_{i1j}$
vanishing. There are three cyclic attractors, namely 
$(123)$, $(243)$ and $(1342)$. 
{\bf (b)} Enumeration of all $N(N-1)=12$ directed
edges, the phase-space elements.
{\bf (c)} The corresponding phase-space graph.
{\bf (d)} The same realization of the routing table
as in {\bf (a)}, now in terms of the phase-space graph.
}
\label{fig_4sites}
\end{figure}

Transport on networks, like the spreading of 
rumors~\cite{Moreno04} and diseases~\cite{klovdahl85} 
in social networks or the flow of capital in financial 
networks~\cite{fronczak11} has been studied intensively,
indeed transport constitutes a basic process in 
biology quite in general~\cite{anderson95}, as well as in
sociology and technical applications. In many cases the
quantity transported is not conserved, e.g.\ when considering
the spreading of rumors in social networks. 
Routing processes, investigated here, model the transport 
of a conserved quantity, 
like conserved information packages. Information 
packages are sent from node to node and are routed at 
every vertex, as illustrated in Fig.~\ref{fig_4sites}. 
A routing process eventually ends up in one of the cyclic 
attractors, the members of the attractors benefiting hence
from a continuous flow of information arriving from the
respective basins of attraction. We have shown previously 
that the geometric arrangement of the attractors on the 
network gives rise in the thermodynamic limit to a non-trivial
distribution for the information centrality, 
which measures the number of attractors intersecting 
at a given vertex~\cite{markovic09}.

We present here the solution for two types
of routing models, Markovian routing in
the absence of a routing memory and vertex routing
in the presence of an one-step memory. The 
solutions are asymptotically exact in the
thermodynamic limit $N\to\infty$, they can be evaluated
for large networks containing thousands to millions of sites.
We present results for the scaling behavior of the 
overall number of attractors and for the mean
of the cycle length distribution. We find, that
the number of cycles increases as $\log(N)$
and that the mean cycle length scales like
$\sqrt{N}/\log(N)$ and $N/\log(N)$ respectively
for the model without and with routing memory.

We also derive rigorous results for the case of
stochastic sampling of phase space, which yields
a cycle length distribution weighted by the size of the
respective basins of attraction. This kind of `on the fly'
sampling is generically equivalent to an experimental
observation of a real-world dynamical system. We find
power law scaling for on-the-fly sampling, logarithmic
corrections are absent. We conclude that
real-world investigations of scaling in complex dynamical 
systems, like the brain, need to be interpreted carefully.

\section{Models}
The two classes of models we consider differ with respect
to the absence/presence of a routing memory. The phase space
volume $\Omega$ is respectively linear and quadratic in the
number of vertices $N$.
\begin{itemize}
\item For the Markovian model the selection of the next active
vertex is independent of the previous state~\cite{meyn93}.
At every point in time only one vertex is active, the vertex
with the information package. The phase space is hence identical
with the collection of vertices; $\Omega = N$;

\item For the vertex routing model the phase space is given by the
collection of directed links; $\Omega=N(N-1)$. At every point in 
time one directed link is active, the link currently transporting 
the information package, compare Fig.~\ref{fig_4sites}. 
\end{itemize}

In both setups the routing of information packages is 
realized through static routing tables. For every
incoming edge the routing table specifies an allowed
outgoing edge.  
A vertex $k$ will transmit an information package, which 
was received from a vertex $j$, to a specific neighboring
vertex $i$. The vertex routing table $\hat{T}$
corresponds to a tensor of binary elements 
$T_{ikj}=(\hat{T})_{ikj} \in \{0,1\}$,
\begin{equation}
T_{ikj}\ =\ \left\{
\begin{array}{rll}
0 &\quad\hbox{no\ routing\ allowed}& \\
1 &\quad\hbox{routing\ from} &
\vec{e}_{jk}
\hbox{\ to\ } \vec{e}_{ki}
\end{array}
            \right. ~,
\label{eq_routing_table}
\end{equation}
where $\vec{e}_{jk}$ denotes a directed edge from vertex 
$j$ to vertex $k$. An example of a routing table
for a four-site network is presented in Fig.~\ref{fig_4sites}. 
In Fig.~\ref{fig_4sites} (a) allowed routing paths are color 
coded and mapped to a four-site network. The complete phase 
space of this network is obtained by representing each 
edge (Fig.~\ref{fig_4sites} (b)) as a node in an iterated 
graph which is shown in Fig.~\ref{fig_4sites} (c). Here each 
node corresponds to a same colored and numbered edge shown in 
Fig.~\ref{fig_4sites} (b). In Fig.~\ref{fig_4sites} (d) 
we show again a single realization of routing tables, but 
now in the iterated phase space graph. The edges of the phase 
space graph shown correspond to allowed routing directions, 
that is, to non-zero entries of the routing table $\hat{T}$.

We consider here critical models, {\it viz.} models
where the number of information packages is conserved.
When the information is received along edge $\vec{e}_{jk}$, 
it can hence be transmitted along only one outgoing 
edge $\vec{e}_{ki}$, 
\begin{equation} 
\label{eq_entries}
\sum_i T_{ikj} = 1, \qquad
\sum_{ij} T_{ikj} = z_k~,
\end{equation}
the non-zero entries of the routing table are drawn randomly. 
Here $z_k$ is the degree of vertex $k$, which is $N-1$ for 
fully connected networks considered here. For the Markovian 
model the routing table $T_{ikj}$ is
independent of $j$, that is, routing depends only on the node 
which received the information package and not on the direction 
along the information was received.

\begin{figure}[t]
\centerline{
\includegraphics[width=0.8\textwidth,angle=0]{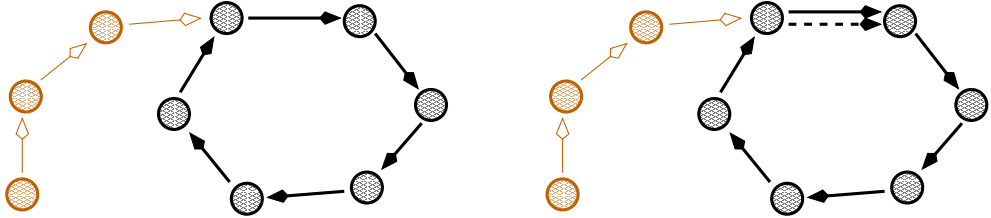}
           }
\caption{Random walks through configuration space for the 
Markovian model (left) and for the vertex routing model (right).
In order to find an attractor independent of the size of their
basins of attraction (light color) one needs to close the path
at the respective starting points. The probability to find a
given attractor is, on the other side, proportional to
the size of its basin of attraction for stochastic `on the fly'
sampling of phase space.
}
\label{fig_random_walks}
\end{figure}

\section{Cycle length distribution}
The dynamics consists of random walks through configuration 
space, as illustrated in Fig.~\ref{fig_random_walks}. One can hence 
adapt the considerations~\cite{gros08}, used for solving the 
Kauffman network for large connectivity $K\to\infty$, in order
to solve the vertex routing model analytically. In addition 
to the previously derived expression for cycle length distribution 
in the case of the Markovian model \cite{markovic09}, we 
present here the solution of the vertex routing model. 

The general expression for the average number of cycles 
$\langle C_L \rangle$ of length $L$ is given by
\begin{equation}
\label{eq_cyclength}
\langle C_L \rangle_N(r)=\frac{N(N-1)^r}{L(N-1)^{r+1}}~q_r(t=L-1),
\end{equation}
where $r=0$ for the Markovian model and $r=1$ for the vertex routing 
model. Here the factor $1/L$ cancels overcounting of a cycle 
of length $L$, while the factor $N(N-1)^r$ is the number of phase 
space elements, that is the number of possible starting elements. 
The factor $1/(N-1)^{r+1}$ gives the probability to close the cycle 
exactly at the starting phase space element. For the Markovian model 
the probability to close the cycle at the starting node is inversely 
proportional to the number of neighbors, whereas in the vertex routing 
model this probability is inversely proportional to the squared 
number of neighbors as the initial edge has to be matched for 
closing the path (see Fig.~\ref{fig_random_walks}). 
The $q_r(t=L-1)$ is the probability that a path containing $L$ 
nodes is still open. At a time step $t = 0,1,\ldots$, we have already 
visited $t$ nodes. Thus, a probability that the next node in 
the sequence was already visited is $t/(N-1)$. For the trajectory
to enter a cycle, the routing has to retrace the existing path. 
The probability for this to happen is $1/(N-1)^r$. The relative 
probability of closing the path at next time step is then 
$\rho_r(t) = t/(N-1)^{r+1}$. This relation constitutes an
approximation, for finite $N<\infty$, in the case of the vertex 
routing model, as self-intersecting paths are neglected.

The probability of still having an open path after 
$t+1$ steps is
\begin{equation}
 q_r(t+1)=q_r(t)(1-\rho_r(t)).
\end{equation}
Expanding the equation till the term $q_r(1) = 1$ and substituting the 
expression for relative probability one obtains 
\begin{equation}
\label{eq_open_probability}
 q_r(t) = \frac{((N-1)^{r+1}-1)!}{(N-1)^{(r+1)(t-1)}((N-1)^{r+1}-t)!}~.
\end{equation}
Substituting (\ref{eq_open_probability}) in (\ref{eq_cyclength})
for the Markovian model, given by $r=0$, one finds
\begin{equation}
\label{eq_C_LN_m}
\langle C_L\rangle_{m}(N) = \frac{N!}{L(N-1)^{L}(N-L)!}
\end{equation}
for the average number of cycles of length $L$.
For the vertex routing model, given by
$r=1$,  the average number of cycles is
\begin{equation}
\label{eq_C_LN_v}
\langle C_L\rangle_{v}(N)=
\frac{N((N-1)^{2})!}{L(N-1)^{2L-1}((N-1)^{2}+1-L)!}~,
\end{equation}
Note that for the Markovian model the cycle length $L$ 
falls within a range $\{2, N\}$, while 
$L \in \{2, (N-1)^2+1\}$ for the vertex routing model.

Relation (\ref{eq_C_LN_v}) 
is an approximation to the average number of cycles as it 
doesn't take into account corrections for self intersecting 
paths. These corrections drop however as ${1}/{N}$ and can 
be neglected in the thermodynamic limit. Furthermore, the 
graph of the phase space elements (see Fig.~\ref{fig_4sites} (c))
is not fully connected and thus not Hamiltonian for arbitrary 
network size $N$, which means that cycle visiting every element 
of the phase space do in general not exist. Formulas (\ref{eq_C_LN_m}) 
and (\ref{eq_C_LN_v}) are based on a mapping to random maps 
and can be generalized to the case of routing on $NK$ networks.

The probability of observing a cycle of length $L$ is obtained 
by dividing the average number of cycles of length 
$L$ from (\ref{eq_C_LN_m}) and (\ref{eq_C_LN_v}) by the 
total number of cycles in a single realization of 
the routing table which is given as
$$
\langle n \rangle_{v,m} = \sum_L\,\langle C_L\rangle_{v,m}~.
$$
We denote with
$$
\rho_{m,v}(L,N), \qquad\quad \sum_L \rho_{m,v}(L,N)=1
$$
the normalized cycle length distributions for the 
Markovian (m) and for the vertex routing model (v), 
Note that substituting $N$ by $(N-1)^2+1$ in (\ref{eq_C_LN_m}) 
one obtains for large N the approximate scaling relation
\begin{equation}
\langle C_L\rangle_v(N) \ \sim\ 
\langle C_L\rangle_m((N-1)^2+1)
\label{eq_scaling}
\end{equation}
between the number of cycles 
of the vertex routing and the Markovian model,
$\langle C_L\rangle_v$ and $\langle C_L\rangle_m$.

\begin{figure}[t]
\centerline{
\includegraphics[width=0.7\textwidth]{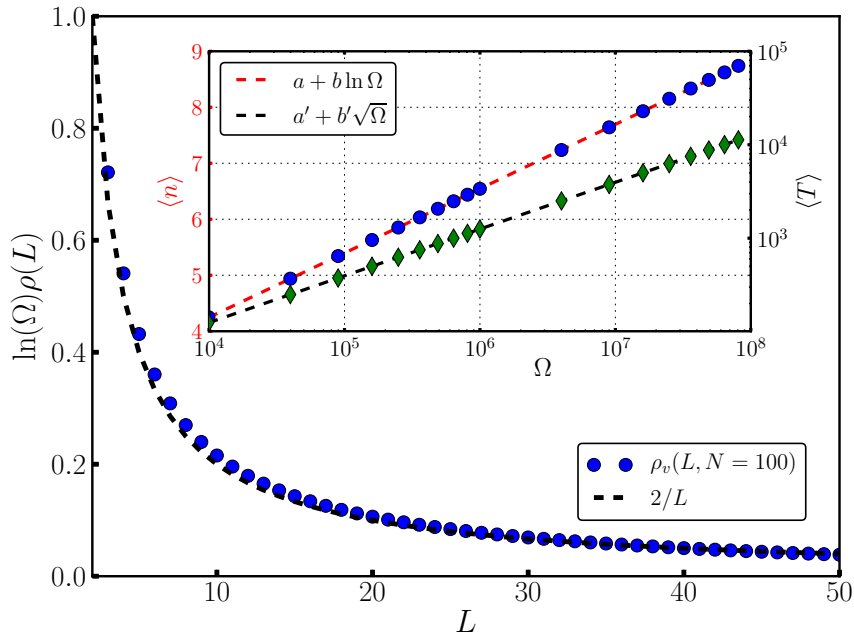}
           }
\caption{The cycle length distributions 
$\rho_{v}(L)$, rescaled by $\log(\Omega)$,
for the vertex routing model. The
dashed line, $2/L$, represents the large-$N$ 
and small-$L$ limiting behavior.
In the inset two quantities are plotted 
as a function of the phase space volume 
$\Omega$.
The average number of cycles $\langle n \rangle$  
(see Eq.~(\ref{eq_n_total}), filled blue circles,
log-linear plot) and the expected total cycle 
length $\langle T \rangle$ (see Eq.~(\ref{eq_T_total}), 
green filled diamonds, log-log plot).
Also included are fits using $a+b\ln\Omega$ 
(red dashed line), with $a = -0.345(3)$ and $b =0.4988(2)$, 
and using $a^{\prime} + b^{\prime}\sqrt{\Omega}$ 
(black dashed line) with $a^{\prime} = -0.3311(5)$ and 
$b^{\prime} = 1.25331 \pm 2\cdot10^{-7}$. 
The coefficient of determination is $R^2 = 1.0$ in both cases, 
within the numerical precision.
}
\label{fig_cycle_length_distribution}
\end{figure}

\begin{table}[b]
\caption{ 
\label{tbl_exponents}
Scaling relations, as a function of the number of vertices 
$N$, for the number of cycles and for the mean of the cycle 
length distribution, respectively for vertex 
routing (v) and the Markovian (m) model. The
routing table distribution is either quenched 
(exact result) or generated on the fly, as
it corresponds to a stochastic sampling of phase
space. Only relative quantities can be 
evaluated for on the fly dynamics.\newline}
\begin{tabular}{|l|l|l|l|}
\cline{3-4}
\multicolumn{2}{c|}{}
 & \ quenched \ & \ on the fly \ \\
\hline
\multirow{2}{*}{\ (v)\ }
    & \ number of cycles\ & \ $\log(N)$         \ & \ -- \ \\
    & \ mean cycle length  \ & \ $N/\log(N)$       \ & \ $N$ \ \\
\hline
\multirow{2}{*}{\ (m)\ }
  & \ number of cycles\ & \ $\log(N)$          \ & \ --       \ \\
  & \ mean cycle length  \ & \ $\sqrt{N}/\log(N)$ \ & \ $\sqrt{N}$ \ \\
\hline
\end{tabular}
\end{table} 

\section{Results}

The analytic expressions (\ref{eq_C_LN_m}) and 
(\ref{eq_C_LN_v}) for the number of attractors
are valid for quenched dynamics \cite{gros08}, 
viz for fixed routing tables. One can,
in addition, evaluate the number of cycles
obtained when randomly sampling phase space,
which corresponds to generating the routing 
tables on the fly. The corresponding results
will be discussed in 
Sect.~\ref{Stochastic_sampling_of_phase space}.

\subsection{Quenched dynamics}
\label{Quenched_dynamics}

Evaluating numerically the number of cycles
(\ref{eq_C_LN_m}) and (\ref{eq_C_LN_v}) we find,
see inset of Fig.~\ref{fig_cycle_length_distribution},
that the total number of attractors 
\begin{equation}
\langle n \rangle_{v,m} = 
\sum_L\,\langle C_L\rangle_{v,m}
\label{eq_n_total}
\end{equation}
growth logarithmically,
as a function of phase space volume $\Omega$.
This result is consistent with a direct evaluation
of the number of attractors for random maps~\cite{kruskal54}. 
The total number of cycles hence grows slower
than any polynomial of the number of vertices $N$,
in contrast to critical Kauffman models, where it
grows faster than any power of $N$~\cite{samuelsson03}.

The normalized cycle length distributions 
$\rho_{v,m}(L) = \langle C_L\rangle_{v,m}/\langle n \rangle_{v,m}$
thus scale as $1/\log(\Omega)$, due to the divisor $\langle n \rangle_{v,m}$. 
The rescaled distributions $\log(\Omega)\rho_{v,m}(L)$ 
approach the thermodynamic limit rapidly, 
compare Fig.~\ref{fig_cycle_length_distribution}.
For small cycle lengths $L$ the limiting functional form
of the rescaled distributions is $2/L$, while for large 
$L \rightarrow L_{max}$ it falls off as 
$(1-L/L_{max})^{(L-L_{max}-1/2)}{\rm e}^{-L}$.
The limiting behavior of 
$\log(\Omega)\rho_{v,m}(L)$ is identical for both models, 
due to the intermodel scaling relation (\ref{eq_scaling}).

The total cycle length, {\it viz.} the combined length of 
all cyclic attractors present for a given system size $N$, 
is on the average
\begin{equation}
\langle T \rangle _{v,m} = \sum_L L\langle C_L\rangle_{v,m}~.
\label{eq_T_total}
\end{equation}
The total cycle length follows a polynomial growth as the function 
of phase space volume $\Omega$ (see the inset of 
Fig.~\ref{fig_cycle_length_distribution}). This
algebraic dependence of the total cycle length 
can be obtained analytically by generalizing the 
analysis~\cite{kruskal54} for the $N\to\infty$ limiting
behavior of the mean cycle length (\ref{eq_n_total}) 
to $\langle T \rangle _{v,m}$.

\begin{figure}[t]
\centerline{
\includegraphics[width=0.7\textwidth]{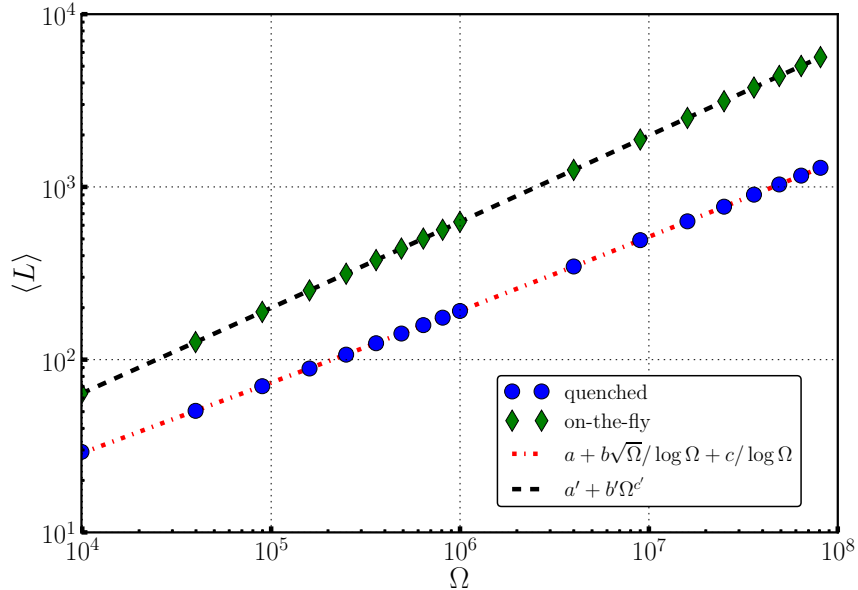}
           }
\caption{Log-log plot, as a function of the
phase space volume $\Omega$, of the mean cycle 
lengths $\langle L\rangle_{v,\tilde{v}}$,
see Eq.~(\ref{eq_mean_length}), 
for the vertex routing with quenched dynamics 
($\langle L\rangle_v$, blue circles) and the vertex routing with 
on the fly dynamics ($\langle L\rangle_{\tilde{v}}$, 
green diamonds). The dashed lines are fits using
$a+b\sqrt{\Omega}/\log(\Omega)+c/\log(\Omega)$ 
and $a^\prime + b^\prime \Omega^{c^{\prime}}$ respectively,
with $a = 8.1(8)$, $b = 2.6035(9)$, $c = -69(9)$, 
and $a^{\prime} = 1.3319(3)$, $b^{\prime} =  0.406(5)$, 
$c^{\prime} = 0.5 \pm 9\cdot10^{-8}$. The coefficient of 
determination is $R^2 = 1.0$ in both cases, within the 
numerical precision. 
}
\label{fig_scaling_mean}
\end{figure}
%

The determination of the scaling behavior is somewhat 
more subtle for the mean cycle length (see
Fig.~\ref{fig_scaling_mean}). 
\begin{equation}
\label{eq_mean_length}
\langle L \rangle_{v,m} = 
\frac{\langle T \rangle _{v,m}} {\langle n \rangle_{v,m}} 
= \sum_L L\rho_{v,m}(L)
\end{equation}
We find that the functional dependence on the phase 
space volume is best reproduced by 
$a+b\sqrt{\Omega}/\log(\Omega)+c/\log(\Omega)$, 
where $a,\, b,\, c$ are free parameters, which fits the data
by about one order of magnitude better than a pure power law
Ansatz $a'+b'\,\Omega^{c'}$. This dependence is obtained 
by keeping the fastest growing terms of mean cycle length 
as $\Omega \rightarrow \infty$. 
Note that $a$, and respectively
$a'$, are finite size corrections not obtainable when
evaluating analytically the scaling of
(\ref{eq_n_total}) and (\ref{eq_T_total}) separately.
Interestingly, log-corrections to power law scaling 
have been studied also in sandpile models at the 
upper critical dimension \cite{lubeck98}
and in epidemic percolation \cite{janssen03}.
An overview of the obtained scaling
relations is given in Table~\ref{tbl_exponents}, where
`quenched dynamics' denotes the results for quenched
distributions of routing tables (exact result).
Note that in Figs.~\ref{fig_cycle_length_distribution} 
and \ref{fig_scaling_mean} we present only the data for
the vertex routing model as it completely overlaps
for large phase spaces $\Omega$, due
to the scaling (\ref{eq_scaling}), with the results
for the Markovian model.
 
\subsection{Stochastic sampling of phase space}
\label{Stochastic_sampling_of_phase space}

In addition to working with predetermined (quenched) 
vertex routing tables one can generate dynamics 
`on the fly' without explicitly creating initially routing 
tables for all vertices of the network. For this 
kind of dynamics, which correspond to a stochastic
sampling of phase space, a routing for a given vertex is 
selected only when the trajectory visits this vertex. 
A cyclic attractor is then found when one state of 
the phase space (edge or node) is visited more then once.
The probability to find a cycle is hence weighted
by the size of its basin of attraction.

The probability of observing a closed cycle of length 
$L$ in a randomly generated path of length $t$ 
after a total number of $t$ routing steps is
\begin{equation}
p(L|~t) = \frac{\Theta(t-L)\Theta(L-2)}{t-1}~,
\end{equation}
where $\Theta(x)$ is the Heaviside step function with 
$\Theta(0)=1$. The joint probability distribution 
$P(L,t)$ is given as $P(L,t)=p(L|~t)p_t$, where 
$p_t = q_t\rho_t$ is the probability of closing 
a cycle at the next time step $t+1$. Then, the 
probability of generating a cycle of length $L$ becomes 
simply the sum over all possible path lengths, with the maximum path 
length $t_\mathrm{max}=N$ for the Markovian routing and 
$(N-1)^{2}+1$ for routing with memory.
Thus, the probability to find an L-cycle is 
$$
\tilde \rho_{v}(L,N)=
\sum_{t=L}^{L_{max}}\frac{((N-1)^{2})!}{(N-1)^{2t}((N-1)^{2}+1-t)!}~,
$$
where we denoted with $\tilde \rho_v(L,N)$ the weighted 
cycle length distribution for the vertex routing model,
viz the cycle length distribution for on-the-fly dynamics.
An analogous relation holds for the Markovian model. 
By generalizing the scaling relation (\ref{eq_scaling}) one finds
$\tilde \rho_v(L,N)=\tilde \rho_m(L,(N-1)^2+1)$ and consequently
$\langle L \rangle_{\tilde{v}}(N)=\langle L \rangle_{\tilde{m}}((N-1)^2+1)$,
where $\tilde \rho_m$ denotes the weighted cycle length distributions 
for the Markovian model.

Fitting the data, as shown in Fig.~\ref{fig_scaling_mean}
for the vertex routing model, with and without log-corrections,
we find evidence for a scaling $\sim{N}$
and $\sim\sqrt{N}$ for the mean cycle lengths of the
vertex routing and the Markovian model respectively with
on-the-fly dynamics. Note that the overall number of cycles 
cannot be obtained when routing on the fly, only relative quantities 
can be evaluated.

\section{Discussion}
For Boolean networks the phase space volume $\Omega$ is
$2^N$ and hence grows exponentially with the number
of vertices $N$. The fact~\cite{samuelsson03},
that the number of attractors grows faster than 
any power of $N$ could in principle be related to the 
exponential growth of the phase space
volume. Our results however show, that the critical
properties of the Kauffman networks for connectivity 
$Z=2$,  and of the vertex routing models considered here 
are not related. The scaling $\sim \log(\Omega)$ valid 
for vertex routing models would imply a polynomial scaling
with the system size  
$$
\log(\Omega) \ \sim\ N, \qquad\quad \Omega=2^N
$$
for critical Kauffman nets, which is however not
observed~\cite{samuelsson03}. Our results hence indicate 
that scaling in critical dynamical systems may generically 
be non-universal, depending on the details of the microscopic
dynamics.

We also note that other properties of critical 
dynamical systems, like the scaling of the number 
of frozen or relevant nodes for critical Boolean 
networks~\cite{mihaljev06}, may show highly 
non-trivial behavior. For the case of vertex routing models 
one may define a measure of centrality, information centrality, 
determined by the number of attractors intersecting a given 
vertex, which scales to a non-trivial limiting distribution 
in the thermodynamic limit~\cite{markovic09}.

Our results may also be seen in the context of
the surge in interests in modelling~\cite{gros09,markovic10} 
and in experimentally investigating~\cite{ringach09,petermann09} 
the spontaneous neural dynamics of the brain.
The observation of power law scaling relations~\cite{eguiluz05} 
has been interpreted as evidence of a critical 
self-organized neural state~\cite{chialvo10}. 
The power law scaling in neural activity was observed 
in spite of strong sub-sampling of neural avalanches 
resulting from small number of electrodes relative to 
total number of neurons within the cortex. Priesemann and 
colleges \cite{priesemann2009} have recently demonstrated 
that sub-sampling of critical avalanches results in the loss 
of power law scaling, thus suggesting different causes of 
the power law scaling of neural avalanches observed in 
various experiments in spite of low number of electrodes, 
used to record neural activity, compared to a total number 
of neurons. 

Our results suggest, to some extent, that 
there is no universal relation in dynamical systems theory between 
criticality and power law scaling and that scaling is generically
dependent on the observation modus. The unbiased statistics
of a certain property, like the number of attractors or avalanches,
may differ from a statistics obtained via stochastic sampling 
($\rho_{v,m}(L)$ and $\tilde\rho_{v,m}(L)$ in our case). The
later will in general be dependent on the size of the respective
basins of attraction of the dynamical process considered, viz
of a cycle or an avalanche. For the case of the vertex routing models
studied here we found logarithmic corrections to power law scaling
for the unbiased, quenched statistics and pure power law scaling 
for stochastic on the fly sampling. We conclude that experimental
observations of real-world systems, when investigating scaling,
 need to be interpreted carefully.


\end{document}